\documentclass[aip,jap,12pt]{revtex4-2}

\usepackage{xcolor}
\usepackage{graphicx}
\usepackage{float}
\usepackage{xr}

\begin{document}

\title{Nickel intercalation in epitaxial graphene on SiC(0001): a novel platform for engineering two-dimensional heterostructures} %Title of paper

\author{Ylea Vlamidis}
\affiliation{Istituto Nanoscienze–CNR, NEST-Scuola Normale Superiore, Piazza San Silvestro 12, 56127 Pisa, Italy}
\affiliation{Department of Physical Science, Earth, and Environment, University of Siena, Via Roma 56, 53100 Siena, Italy}

\author{Stiven Forti}
\affiliation{Center for Nanotechnology Innovation@NEST, Istituto Italiano di Tecnologia, Piazza San Silvestro 12, Pisa 56127, Italy}

\author{Antonio Rossi}
\affiliation{Center for Nanotechnology Innovation@NEST, Istituto Italiano di Tecnologia, Piazza San Silvestro 12, Pisa 56127, Italy}
\affiliation{Graphene Laboratories, Istituto Italiano di Tecnologia, Genova 16163, Italy}

\author{Arrigo Calzolari}
\affiliation{Istituto Nanoscienze–CNR, S3, Via Campi 213/a, Modena 40125, Italy}

\author{Carmela Marinelli}
\affiliation{Department of Physical Science, Earth, and Environment, University of Siena, Via Roma 56, 53100 Siena, Italy}

\author{Camilla Coletti}
\affiliation{Center for Nanotechnology Innovation@NEST, Istituto Italiano di Tecnologia, Piazza San Silvestro 12, Pisa 56127, Italy}
\affiliation{Graphene Laboratories, Istituto Italiano di Tecnologia, Genova 16163, Italy}

\author{Stefan Heun}
\affiliation{Istituto Nanoscienze–CNR, NEST-Scuola Normale Superiore, Piazza San Silvestro 12, 56127 Pisa, Italy}

\author{Stefano Veronesi}
\affiliation{Istituto Nanoscienze–CNR, NEST-Scuola Normale Superiore, Piazza San Silvestro 12, 56127 Pisa, Italy}

\date{\today}

\begin{abstract}
\section*{Abstract}
Two-dimensional (2D) magnetic materials integrated with graphene offer a compelling platform for next-generation spintronic devices, yet nickel in its 2D form remains largely unexplored, due to fundamental synthesis limitations. Here, we report the controlled intercalation of Ni beneath epitaxial graphene on the Si-face of SiC(0001), achieved through a scalable colloidal nanoparticle deposition route. Chemically synthesized Ni nanoparticles ($\sim$10 nm diameter) are uniformly deposited onto graphene via immersion in colloidal solution at room temperature; subsequent thermal annealing at 650 °C drives intercalation, yielding well-ordered Ni islands at the graphene/buffer-layer interface with morphology dictated by annealing conditions. Scanning tunneling microscopy (STM) and angle-resolved photoemission spectroscopy (ARPES), supported by density functional theory (DFT) calculations, elucidate the atomic and electronic structure of the intercalated layers. DFT simulations further confirm the thermodynamic stability of the 2D nanostructures as a function of shape and lateral size, predicting a robust average magnetic moment of ~0.9 $\mu_B$ per atom. The resulting Ni-intercalated graphene on SiC constitutes a well-defined 2D heterostructure combining preserved graphene band structure with robust interfacial magnetism, stable under ambient conditions. These findings establish a reproducible, scalable pathway to engineer magnetic graphene-based heterostructures and open new avenues for their integration into spintronic architectures.

\end{abstract}

\maketitle

\section{Introduction}

Spintronics exploits the intrinsic spin of electrons to enable ultra-fast processing, high-density data storage, and low-power operation,\citep{Puebla2020, Li2019} making it a key technology for next-generation electronics such as magnetic sensors, logic devices, magnetoresistive random access memory (MRAM), and quantum computing components.\citep{Gloag2019, Incorvia2024, Zutic2004} Advancing spintronic applications requires materials that enable efficient spin transport and stable spin textures.\citep{Zhou2024}

This has spurred significant interest in two-dimensional (2D) magnetic materials, which combine long-range magnetic order at the monolayer level with unique properties such as mechanical strength and flexibility, optical transparency, tunable surface chemistry, and facile integration into heterostructures without lattice mismatch issues.\citep{SanchezYamagishi2025, Khan2020, Lei2022} 2D-spintronics enables unique functionalities, such as imprinted skyrmions which exist only at 2D interface. Moreover, neuromorphic computing requires scaling down the structure dimensions for obtaining ultra-high density integration. This target is difficult to achieve with traditional 3D materials, whereas the availability of 2D magnets provides a promising solution.\citep{Zhang2024, Papavasileiou2023} 2D magnetic materials such as CrI$_3$,\citep{Fu2024} Fe$_3$GeTe$_2$,\citep{Gong2019} and MnBi$_2$Te$_4$ \citep{Wang2018} exhibit long-range magnetic order, with Fe$_3$GaTe$_2$ being the only one to show deterministic magnetic switching at room temperature.\citep{Kajale2024, Wang2019, Kratochvilova2022} However, spintronic performance in such systems strongly depends on interface quality, which is often compromised by defect oxidation and degradation of these ferromagnetic materials in air.\citep{Wang2019a, Shcherbakov2018} Magnetic van der Waals materials can be isolated in 2D form via mechanical exfoliation, however, like all exfoliated materials, they inherently suffer from poor scalability.\citep{Zhang2024, Papavasileiou2023} Thus, fabrication assumes a critical role in the magnetic response of the samples in the quest to realize robust, magnetically functional 2D materials compatible with device integration. Indeed, integrating 2D magnetic materials into device manufacturing requires wafer-scale growth to enable scalability. 

Confinement heteroepitaxy (CHet), in which foreign atoms are intercalated at buried interfaces, has emerged as a powerful strategy to tailor the electronic, structural, and chemical properties of low-dimensional materials.\cite{Yang2024, Briggs2020, Lu2026} Among the explored approaches, the intercalation of metals and other elements at the graphene–SiC interface offers a powerful strategy for modulating key properties such as doping level, band structure, carrier mobility, and local bonding configuration.\cite{Briggs2019, AlBalushi2016,Wundrack2025} This technique enables precise electronic, magnetic, and chemical modifications of graphene while preserving its structural integrity. CHet has been achieved with various atoms via intercalation at the interface between epitaxial graphene (EG) and SiC under ultra-high vacuum conditions.\cite{Forti2020, Fiori2017, GrubisicCabo2024, Rybkina2021, Ferbel2025, Idczak2023}
The epitaxial growth of graphene on SiC proceeds through the sublimation of Si atoms and the subsequent formation of a carbon-rich interfacial layer, commonly referred to as the buffer layer or zero-layer graphene (ZLG). Although the ZLG is topographically indistinguishable from monolayer graphene, approximately one-third of its carbon atoms are covalently bonded to silicon atoms in the substrate, thereby suppressing the formation of the characteristic graphene $\pi$-bands. Owing to its interaction with the SiC(0001) surface and the associated lattice mismatch, the ZLG exhibits a $(6\sqrt{3} \times 6\sqrt{3})R30^\circ$ reconstruction (hereafter referred to as $6\sqrt{3}$).\citep{Riedl2010,Lauffer2008}
Indeed, EG on SiC provides a particularly robust platform due to its large-area, single-crystalline nature and its ability to accommodate intercalated species between the graphene and the underlying SiC substrate. SiC enables the growth of high-quality, single-crystal monolayer graphene with low defect density and excellent structural and electronic properties.\citep{Emtsev2009} Moreover, SiC is especially well suited for exploring interface-confined systems, offering structural stability and technological compatibility.
 Using orbit-orbit coupling,\citep{Kane2005} graphene can sustain long spin coherence lengths, up to micrometer scale spin diffusion at room temperature, making it ideal for spin transport.\citep{Droegeler2016, Choudhuri2019} Magnetic behavior in graphene and other 2D materials can be induced through several approaches, including intercalation.\citep{GonzalezHerrero2016, Giesbers2013, Castro2007, Iqbal2018, MunizCano2024, Hallal2021, Wang2015, Wei2016} 
The intercalation of 2D magnetic layers beneath EG on SiC would provide a platform in which proximity-induced magnetism in graphene is coupled to a confined ferromagnetic phase, enabling controlled spin injection, and transport within a structurally protected and chemically stable environment.

Among possible candidates, nickel stands out for its strong interaction with graphene and its ferromagnetic character. Nickel intercalation has previously been realized in chemical vapor deposition (CVD)-grown graphene on metal substrates such as Ir(0001), Ru(0001), and Rh(0001)\citep{Sicot2010, Huang2011, Gall2000}, and on 6H-SiC(0001).\citep{Robbie2001} However, nickel intercalation has not yet been reported in epitaxial graphene on SiC, representing a critical gap. Nickel possesses a relatively high Curie temperature (354 °C),\citep{Legendre2011} ensuring robust magnetic properties at room temperature. This makes Ni highly appealing for spintronic applications relying on interface magnetism, including spin injection, magnetic proximity effects, and spin filtering.\citep{Mandal2012, Dlubak2012} Embedding Ni layers beneath graphene while preserving their magnetic order offers a promising route toward magnetically functional interfaces for future device architectures, potentially reducing power consumption and improving heat dissipation.\citep{Chen2023} Moreover, further optimization of the nickel intercalation approach could enable wafer-scale scalability, offering attractive prospects for device fabrication. 

In this work, we report the first successful demonstration of nickel intercalation in epitaxial graphene grown on the Si-face of SiC(0001). Ni nanoparticles synthesized via wet chemistry were deposited on the graphene surface and subsequently annealed at temperatures above 650 °C to induce intercalation. Scanning tunneling microscopy (STM) reveals the formation of well-defined monolayer-high geometrical Ni islands at the ZLG-MLG interface. Their edges are aligned with the quasi-($6\times6$) superlattice periodicity of graphene, observed in STM, which arises from the buffer layer and reflects the underlying $6\sqrt{3}$ reconstruction at the SiC interface.\citep{Riedl2007,Rutter2007}
Density functional theory (DFT) provides further information on the local atomic arrangement, electronic structure, and spin interactions, supporting the experimental findings and elucidating the nature of the Ni–graphene interaction. Furthermore, DFT predicts a robust magnetic moment of ~0.9 $\mu_B$ per atom. Moreover, the continuous graphene overlayer acts as a passivation barrier, protecting the confined 2D nickel layer from exposure to air. This environmental stability, combined with the structural precision of the intercalated domains, establishes epitaxial graphene on SiC as a versatile platform for investigating interface-driven phenomena and magnetically functional low-dimensional systems. The successful intercalation of nickel in this system represents a significant step toward the integration of two-dimensional ferromagnetic layers into graphene-based heterostructures, paving the way for future advances in graphene-based spintronic technologies.

\section{Results and Discussion}

In this study, we used epitaxial graphene (EG) samples with a buffer layer (ZLG)/ mono-layer graphene (MLG) ratio of 60:40. The quality, uniformity, and composition of the graphene was initially assessed by Raman spectroscopy and atomic force microscopy (see Figures S1 and S2 in the supporting information for further details).

\subsection{\textit{Surface topography of 2D Ni intercalated in EG}}

Ni nanoparticles (NPs), synthesized with an average diameter of 11 nm, are deposited on graphene at room temperature, where they initially form clusters on the surface. Upon annealing at 550 °C, these clusters remain structurally stable on graphene, as previously reported.\citep{Vlamidis2024} However, at this temperature the increased mobility of the nanoparticles leads to enhanced aggregation, with particles preferentially migrating and accumulating at morphological defects. In particular, they tend to rearrange on the buffer layer, where the surface corrugation is more pronounced, and accumulate along step edges, driven by asymmetric lateral diffusion associated with the Ehrlich–Schwöbel barrier.\citep{Ehrlich1966, Schwoebel1966,Schwoebel1969} As a result of this migration, the monolayer graphene surface becomes largely depleted of large-sized nanoparticles. \citep{Vlamidis2024}

Upon annealing at 650 °C, Ni diffusion and intercalation take place, inducing substantial modifications in the sample morphology as schematically depicted in Figure 1a. In MLG regions, the annealing process leads to the formation of well-defined yet randomly distributed intercalated areas within the terraces, as shown in Figure 1b. In contrast, no evidence of nickel intercalation is observed in the ZLG regions. Large agglomerates, corresponding to Ni clusters, remain visible on the surface of both the ZLG and MLG areas (see Figure 1b and Figure S3), whereas at 650 °C, the small size of the synthesized NPs facilitates atomic intercalation, as thermal activation overcomes interatomic cohesive forces and the associated energy barrier, promoting diffusion beneath the graphene and the formation of a face-centered cubic (FCC) phase at the ZLG–MLG interface. \\
STM scans of areas ranging from $70\times 70$ nm$^2$ and below reveal hexagonal or truncated triangular shaped Ni islands (lateral size $\sim$4-5 nm) as depicted in Figures \ref{fig:1}b and c.
On top of these intercalated islands, the $(6\sqrt{3} \times 6\sqrt{3}) R30^{\circ}$ superstructure of the underlying SiC substrate is visible (see Figures \ref{fig:1}d and e), suggesting that the Ni islands are confined within the van der Waals gap at the  graphene/ZLG interface. This is consistent with the absence of intercalation in ZLG areas, as the partial covalent bonding of the ZLG to the Si substrate makes Ni intercalation beneath it energetically less favorable than beneath the MLG.
Atomic-resolution STM images confirm that the islands are covered by a continuous hexagonal lattice characteristic of graphene (Figure \ref{fig:1}e). Atomic-resolution imaging performed at the edges of the islands confirms that Ni resides beneath the carbon lattice and demonstrates that the intercalation process preserves the structural integrity of the graphene layer. The preservation of the graphene lattice is further supported by the stability of the Ni islands after exposure to ambient conditions and subsequent reintroduction into UHV (see Figure S4). This behavior highlights the effectiveness of epitaxial graphene as a passivation layer, ensuring the long-term stability of the Ni-intercalated structures.

\begin{figure}[htb]
    \begin{center}
    \includegraphics[width=1\textwidth]{Fig_STM_3.jpg}
    \end{center}
    \caption{(a) Sketch of the investigated system before and after annealing. (b-e) STM images of Ni islands intercalated in EG on 6H-SiC(0001) after annealing at 650°C for 2 minutes. (b) 70 $\times$ 70 nm$^2$ STM image showing geometrically defined intercalated Ni islands (yellow circle), and Ni nanoparticle clusters adsorbed on the monolayer graphene region. (c) STM scan over a 50 $\times$ 50 nm$^2$ area highlighting the spatial distribution of the intercalated islands. (d) Magnified 20 $\times$ 20 nm$^2$ view of a single Ni island.  (e) High-resolution 5 $\times$ 5 nm$^2$ image corresponding to the white box in panel (d), revealing both the periodic modulation (dashed rhombus) and the hexagonal graphene lattice on the 2D Ni island. The dashed rhombus highlights the quasi-(6$\times$6) periodic superstructure observed on the surface. Imaging parameters: (b, c) V$_s$ = 0.55 V, I$_t$ = 0.50 nA; (d) V$_s$ =0.35 V, I$_t$ = 0.30 nA; (e) V$_s$ = 0.30 V, I$_t$ = 0.28 nA.}
    \label{fig:1}
\end{figure}

\begin{figure}[htb]
    \begin{center}
    \includegraphics[width=0.8\textwidth]{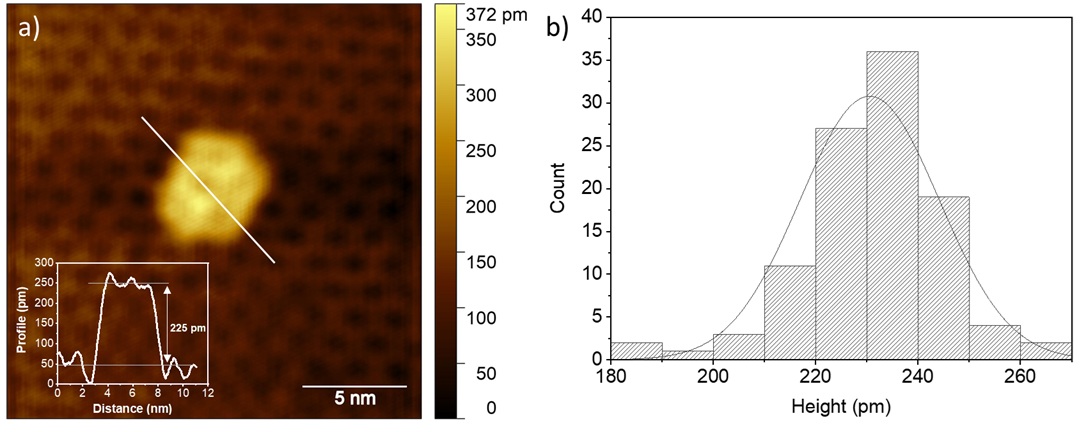}
    \end{center}
    \caption{(a) STM topography of an intercalation region (20 $\times$ 20 nm$^2$), and corresponding height profile (inset) taken along the white line. Imaging parameters: V$_s$ = 0.3 V, I$_t$ = 0.3 nA.(b) Height distribution of intercalated islands obtained from line profiles extracted from STM scans and based on the analysis of 105 individual islands.}
    \label{fig:2}
\end{figure}

Figure \ref{fig:2}a shows a representative Ni-intercalated island with an apparent height of $\sim 225$. The inset in Figure \ref{fig:2}a displays the corresponding line profile acquired across a single island, used to extract its local height. The average island height is measured to be (230.5 ± 13.6) pm, as determined from line profiles collected over a dataset of 105 islands, ensuring statistical reliability of the measurement. Figure \ref{fig:2}b reports the corresponding height distribution histogram obtained from the full dataset of measurements.
The topographic corrugation within the intercalated region is approximately 22 pm, indicating subtle height modulation across the island surface. The average height is close to but even smaller than the step height of a close packed crystal of Ni, indicating that these Ni islands are composed of a single atomic layer. We further estimated how much the intercalant expanded the ZLG-MLG van der Waals gap. In the literature, the typical interlayer distance between the substrate and the buffer layer (223 pm) and between the buffer layer and the first graphene layer (359 pm) are reported.\citep{Emery2013} Considering that Ni intercalates between the buffer layer and the monolayer graphene, the measured distance  between the buffer layer and the lifted graphene layer in the intercalated areas corresponds to $\sim$590 pm.

\subsection{\textit{Orientation and growth}}

Large-area STM topographies reveal that, taking into account the crystallographic orientation of the SiC(0001) terraces, the edges of the Ni islands align along the ⟨1$\overline{1}$00⟩ direction, corresponding to the zigzag orientation of graphene (see Figure 3a).\citep{Riedl2010} This provides clear evidence of an epitaxial registry with the graphene lattice (see also Figure S5), indicating that graphene imposes its crystallographic template on the Ni islands, consistent with the small lattice mismatch ($\sim$ 1.2\%).\citep{Jin2013}

\begin{figure}[htb]
    \begin{center}
    \includegraphics[width=1.0\textwidth]{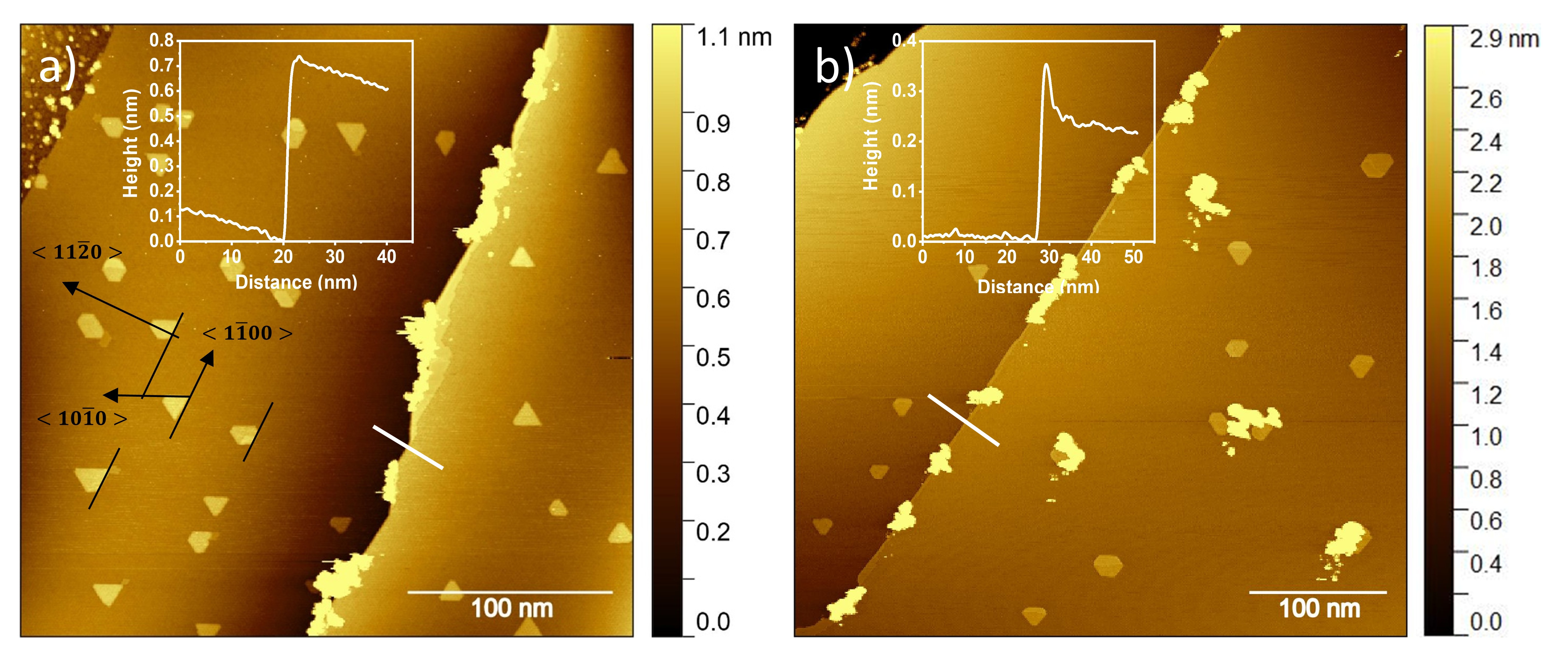}
    \end{center}
    \caption{STM topographic images showing the orientation of Ni islands intercalated in epitaxial graphene (EG). Insets display the corresponding step-edge line profiles for (a) a triple SiC bilayer step, with the in-plane crystallographic directions of the SiC substrate indicated, and (b) a single SiC bilayer step. Contrast-saturated regions correspond to clusters of Ni nanoparticles on the surface. Imaging parameters: (a,b) V$_s$ = 0.7 V, I$_t$ = 0.4 nA.}
    \label{fig:3}
\end{figure}

Despite this local epitaxial alignment, the islands do not adopt random orientations but instead exhibit a terrace-dependent ordering, such that all islands within a given terrace share the same orientation. As shown in Figure \ref{fig:3}a, when the step height reaches three SiC bilayers ($\sim$ 0.750 nm), corresponding to half of the unit cell, islands on adjacent terraces are rotated by 60°. In contrast, this orientation is preserved across terraces separated by either single ($\sim$ 0.250 nm) or double SiC bilayer steps, as shown in Figure \ref{fig:3}b. This behavior reflects the threefold rotational symmetry of 6H-SiC, with a 60° rotation occurring every three bilayers, consistent with the alternating C and Si terminations in the stacking sequence.\citep{MomeniPakdehi2020}
This indicates that, while graphene governs the local crystallographic registry, the preferred Ni island growth directions are influenced by the underlying SiC substrate. Accordingly, the system can be described as epitaxy on graphene with SiC-guided orientational selection, where terrace-dependent symmetry constraints imposed by the substrate modulate the island orientation, evidencing a residual substrate influence despite the presence of graphene as an intermediate layer.
Related substrate-mediated ordering effects have been reported for Ni intercalation at the graphene/Ru(0001) interface, where the buried metal substrate can modulate the structure of intercalated species.\citep{Huang2011}

\subsection{\textit{STM study of the initial stage of Ni intercalation }}

\begin{figure}[htb]
    \begin{center}
    \includegraphics[width=1.0\textwidth]{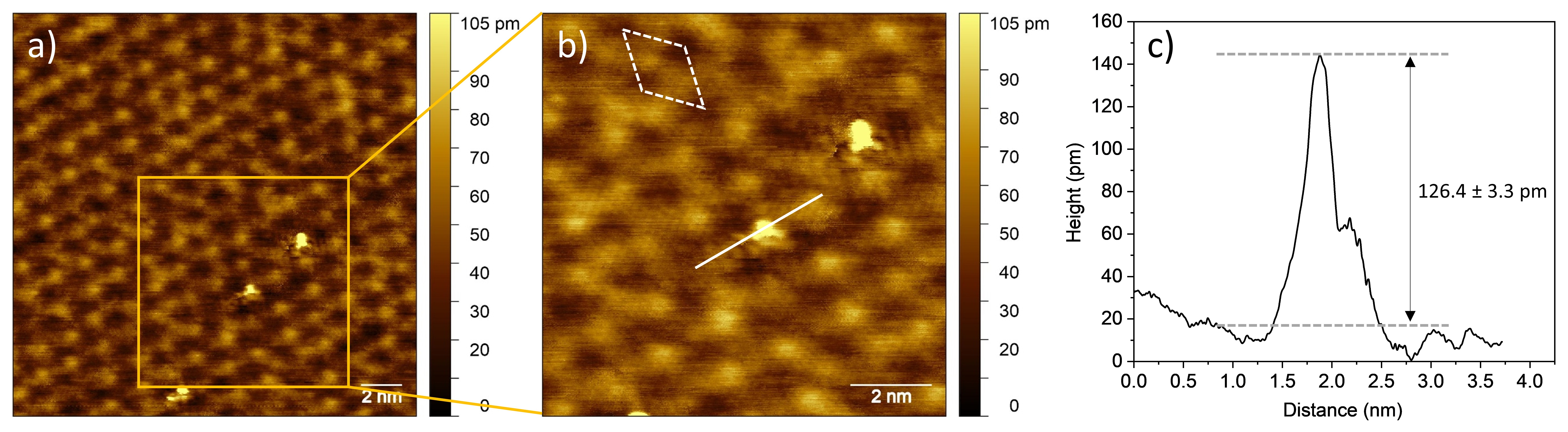}
    \end{center}
    \caption{STM topography images at different magnification showing charge interference patterns around Ni adatoms chemisorbed on the graphene lattice (a: 20 $\times$ 20 nm$^2$ and b: 10 $\times$ 10 nm$^2$), with (b) corresponding to a magnified view of the region indicated by the yellow square in (a). The white dashed rhombus highlights the quasi-(6 $\times$ 6) periodicity of the $6\sqrt{3}$-reconstructed SiC0001. (c) Line profile along the white line in (b), highlighting the atomic corrugation and apparent out-of-plane height of the Ni heteroatom. Imaging parameters: (a) V$_s$= 0.7 V, I$_t$ = 0.60 nA and (b) V$_s$= 0.5 V, I$_t$ = 0.50 nA.}
    \label{fig:4}
\end{figure}

STM was also employed to analyze charge interference patterns generated by electron scattering from individual adatoms on the graphene surface, arising from localized changes in the electronic states induced by heteroadatom doping. Figures~\ref{fig:4}a and b show bright features that we attribute to mobile Ni adatoms adsorbed on the graphene lattice. As previously reported for substitutional nitrogen doping, STM imaging typically reveals a triangular cluster of bright spots, attributed to charge transfer from the dopant to neighboring carbon atoms and the resulting enhancement of the local density of states (LDOS) near the Fermi level.\citep{Sun2023, Zheng2010} Similarly, in our study, Ni doping caused a distinct perturbation in the electronic structure of graphene, evident from the increased brightness surrounding the dopant site. High-resolution STM images (Figure \ref{fig:4}) revealed a triangular arrangement of three bright spots, indicative of a substitutional Ni atom embedded within the graphene lattice. The enhanced contrast stems from electron delocalization from the Ni dopant to the adjacent C atoms, locally increasing the LDOS and providing a clear fingerprint of substitutional Ni incorporation. This electronic perturbation is spatially confined within a few lattice spacings from the dopant site. Additionally, a line scan across the Ni site (Figure \ref{fig:4}c) revealed an out-of-plane displacement of $\sim$1.26 Å, consistent with a Ni atom substituting for a carbon atom within the graphene lattice.\cite{Li2001} The single Ni site appeared as a bright central feature, with surrounding C atoms also exhibiting increased brightness relative to those further from the dopant. This contrast, distinct from the pristine graphene lattice, highlights the strong local modification of the electronic environment caused by substitutional Ni doping.

These STM results also hint at a possible intercalation pathway governed by surface diffusion kinetics. The  mobility of Ni adatoms across the graphene surface implies a low diffusion barrier, which facilitates their migration toward defect sites such as vacancies or step edges, energetically favorable entry points for intercalation. The strong local perturbation of the electronic structure around substitutional Ni sites highlights a strong interaction between Ni and the graphene lattice and indicates that once Ni atoms anchor at such reactive locations, they may find pathways to intercalate into the graphene/ZLG interface. These findings suggest a likely surface-diffusion-mediated mechanism for intercalation, with the graphene overlayer remaining intact and guiding the process through selective, energetically accessible sites.
STM observations indicate that Ni intercalation occurs predominantly within the terraces. Before annealing at 650 °C, surface topographies of monolayer graphene areas show only a few isolated Ni nanoparticles on the terraces. After annealing, intercalated Ni islands appear homogeneously distributed across the terraces. This behavior suggests that, at ~650 °C, the increased mobility of Ni atoms enables diffusion away from the step edges and across the terraces. This is in agreement with what already reported in the literature concerning the intercalation of Ni in graphene on Ru(0001), where the intercalation pathway was suggested to occur by the diffusion of Ni atoms even in the absence of extended defects, due to strong interaction between Ni and C atoms.\citep{Jin2013} The authors found that Ni adsorbed atoms anchoring on the graphene produce transient atomic-scale defects at the graphene lattice, by which Ni can diffuse through the carbon layer. After the Ni atoms bond to the Ru surface, the transient defects are healed, and a perfect graphene surface is reformed. Although Ni atoms arrange in clusters on graphene at elevated temperatures, the Ni intercalation is expected to occur via single-atom penetration, which has a lower barrier compared to the penetration of Ni clusters. A similar mechanism was observed for Si intercalation in analogous systems.\citep{Cui2012}

\subsection{\textit{Effect of the annealing duration on the Ni intercalated islands}}

\begin{figure}[htbp]
    \begin{center}
    \includegraphics[width=0.8\textwidth]{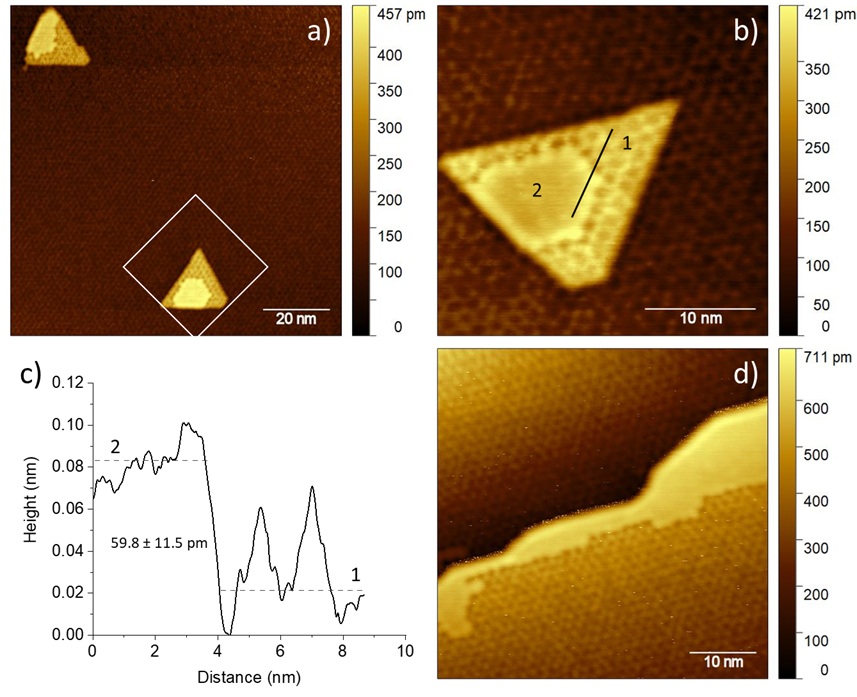}
    \end{center}
    \caption{STM topographies of the EG sample surface after 10 minutes annealing at 650 °C showing the partial formation of a second layer of Ni atoms. (a) STM scan of a 100 $\times$ 100 nm$^2$ area and (b) higher-resolution image (30 $\times$ 30 nm$^2$) of the region indicated by the white square in panel (a). (c) Height profile along the black line in panel b, showing a height difference of $(59.8 \pm 11.5)$ pm between intercalated levels. (d) Ni intercalated areas along a step edge (image size 50 $\times$ 50 nm$^2$). Imaging parameters: (a) V$_s$= 0.4 V, I$_t$ = 0.3 nA; (b) V$_s$= 0.3 V, I$_t$ = 0.28 nA; (d) V$_s$= 0.3 V, I$_t$ = 0.3 nA.}
    \label{fig:5}
\end{figure}

By means of STM measurements, we investigated how  the duration of the annealing treatment and the annealing temperature affect the size and morphology of the intercalated Ni islands. Upon annealing between 650 and 670°C for 2 minutes, we observed islands with average lateral size around 3-5 nm. After a second annealing step at the same temperature for further 3 minutes, the islands grow in lateral size, reaching between 15 and 20 nm (see Figure S6). A further annealing step conducted for 5 more minutes led to the partial formation of a second layer of Ni atoms, which grows starting from the edges of the islands, as shown in Figures \ref{fig:5} a and b. The bilayer Ni regions display a very flat surface, with an rms roughness of 3.3 $\pm$ 1.7 pm, while the step height between mono- and bilayer areas is 59.8 $\pm$ 11.5 pm, as extracted from the height profile in Figure 5c. In regions where a second Ni layer forms, the characteristic $6\sqrt{3}$ surface reconstruction is not entirely suppressed but becomes attenuated (see Figure S7), suggesting that the second metallic layer is also intercalated between the ZLG and MLG. Besides, Ni intercalated areas start to appear uniformly along the step edges (see Figure \ref{fig:5}d).

Conversely, annealing the sample at higher temperature ($\sim$800 °C) results in an irreversible deintercalation of Ni atoms (see Figure S8). This hight-temperature treatment significantly reduces the coverage of intercalated areas, while Ni NP clusters are still present on the graphene surface. Moreover, subsequent annealing around 650°C does not recover the intercalated regions, indicating a permanent alteration in the system, which will be further discussed in the following.

\subsection{\textit{Chemical composition and electronic structure of 2D Ni intercalated in MLG}}

\begin{figure}[htb]
    \begin{center}
    \includegraphics[width=1.0\textwidth]{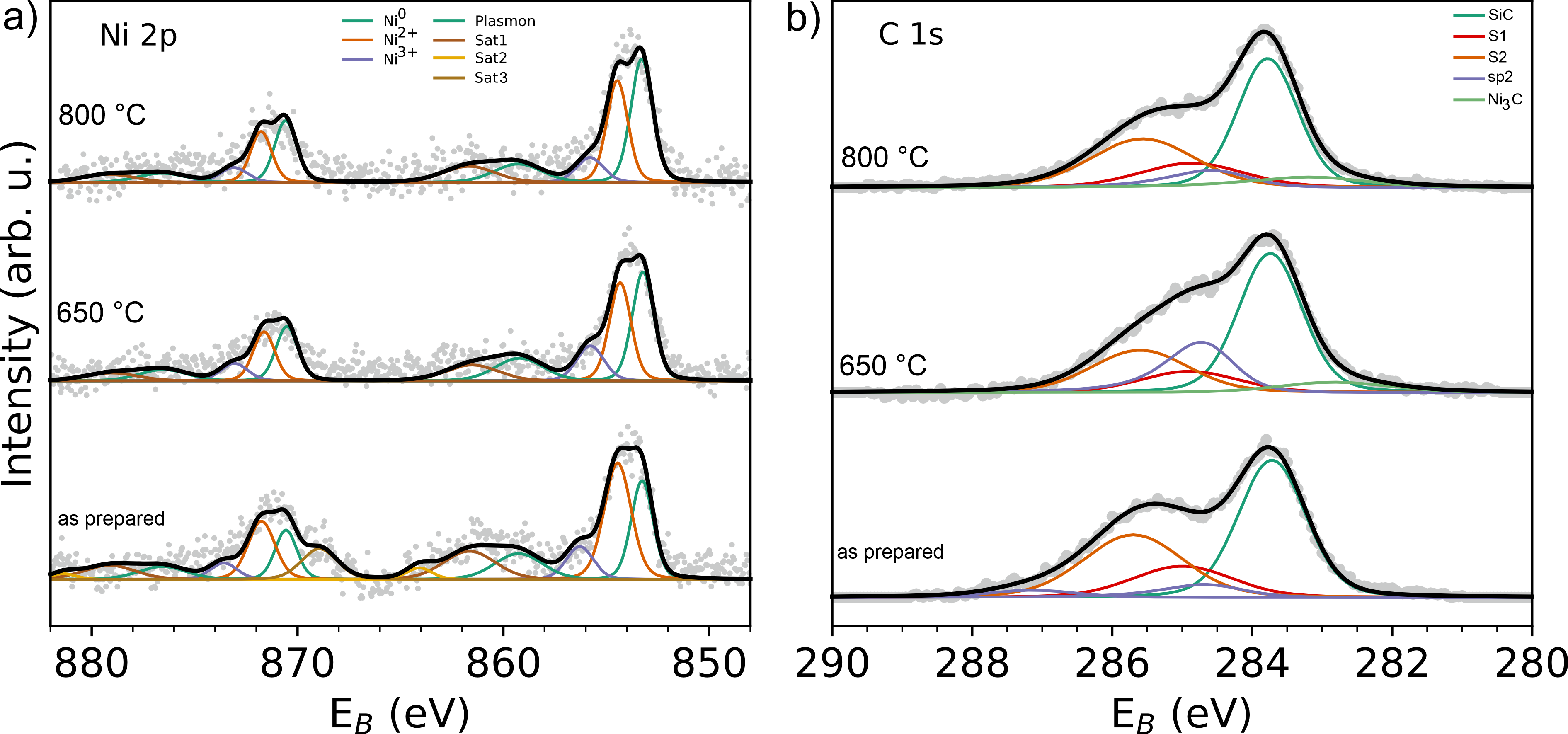}
    \end{center}
    \caption{ Deconvolution of high-resolution XPS spectra of Ni on EG on 6H-SiC(0001) acquired from the as-prepared sample and after subsequent annealing to 650°C and 800°C for: a) Ni 2p, b) C 1s. Experimental data (dots) are shown together with fitted curves (solid lines).} 
    \label{fig:6}
\end{figure}

XPS measurements, conducted to investigate the chemical state and surface composition of the sample following Ni intercalation, support the evidence of Ni intercalation between the ZLG and MLG. The sample was analyzed both as-prepared and after successive annealing at 650 °C and 800°C. Figure \ref{fig:6} presents XPS spectra for the Ni 2p and C 1s core levels. In the Ni 2p spectra (Figure \ref{fig:6}a), three main doublets are observed. The first feature, with its high-spin orbit component at 852.8 eV (Ni 2p$_{3/2}$), is attributed to metallic Ni$^0$. Two additional components were observed: one at 855.8 eV, assigned to Ni$^{2+}$ species (typically nickel oxides\citep{Grosvenor2006,HagelinWeaver2004}, and nickel carbide\citep{Kang2015}), and another at 863.0 eV, corresponding to Ni$^{3+}$ species, which are associated with nickel hydroxides.\citep{Bayer2016, Jia2021, Ajmal2024} After annealing at 650 °C and 800 °C, the Ni\textsuperscript{3+} component decreases in intensity, while the metallic Ni contribution increases.

The C 1s core-level spectra (Figure \ref{fig:6}b) reveals a peak at $\sim$283.7 eV, corresponding to the SiC substrate. Peaks at 284.9 eV and 285.6 eV are attributed to the S1 and S2 components, which reflect the interaction between the interfacial ZLG and the substrate, as reported in literature.\citep{Emtsev2008} For the as-prepared sample, the graphene sp$^2$ signal is weak, likely due to shielding by the randomly distributed CTAB nanoparticle shell, as reported in our previous work.\citep{Vlamidis2024} Following annealing, the SiC component shows no significant variation compared to its room temperature state, further confirming that the SiC interface remains stable and that nickel intercalation between the buffer layer and the SiC is minimal. On the other hand, the contribution to the sp\textsuperscript{2} component increases upon annealing at 650 °C (see also Figure S9a). In regions where Ni-intercalated areas are present, the signal originating from the underlying layers is weakened due to the short attenuation length in Ni. Moreover, the presence of Ni islands at the interface between EG and ZLG increases the electronic screening of the S2 atoms, thereby lowering their measured binding energy. The result is the apparent increase of the sp$^2$ component. In addition, a little shoulder appears on the lower-binding energy side of the SiC component. We ascribe it to the formation of nickel carbide, with Ni atoms binding covalently to the carbon atoms of the buffer layer, which is known for being chemically active. This picture confirms that nickel islands only intercalate in between the mono- and buffer layer graphene on the SiC(0001) surface. At 800 °C, the intensity of the sp\textsuperscript{2} component is restored to its initial intensity, while the nickel carbide component slightly increases, consistent with the deintercalation of Ni atoms observed by STM measurements.

The data collected on the Si 2p core level further corroborate this, as we did not find a signal related to nickel silicide (see Figure S9b).
The intercalation and deintercalation of Ni after annealing steps is further confirmed by ARPES measurements (see Figure S10).

\subsection{\textit{DFT calculations}}

\begin{figure}[htb]
    \begin{center}
    \includegraphics[width=1\textwidth]{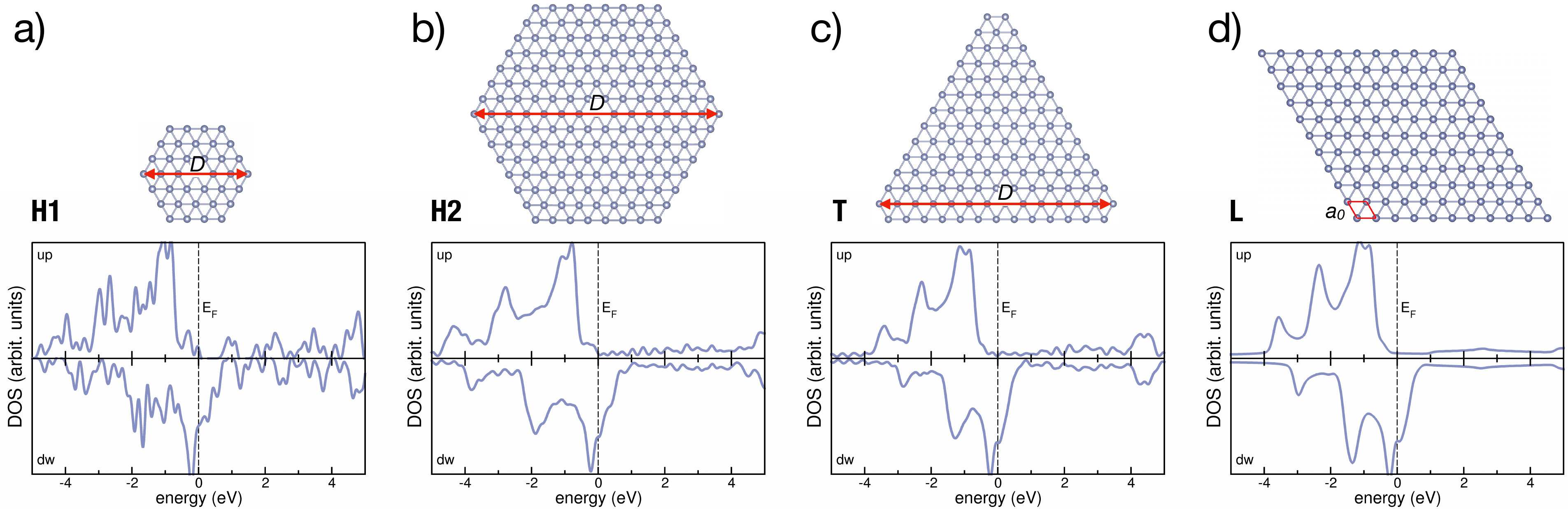}
    \end{center}
    \caption{Atomic structure (top panel) and spin-resolved density of states (DOS, bottom panel) of free standing 2D Ni nanostructures: a) smaller hexagon {\bf H1}, b) larger hexagon {\bf H2}, c) triangular {\bf T} nanoparticles;  d) extended layer {\bf L}. Red lines mark the largest lateral dimension ($D$) of nanoparticles or the lattice parameter ($a_0$) of the 2D layer.Vertical dashed lines in the bottom panel identify the Fermi level (E$_F$) of each system.}
    \label{fig:7}
\end{figure}

To explore the morphology behind the nanostructure growth, we investigated the 
formation of intercalated 2D Ni nanostructures by using first principles simulations based on DFT. We first considered the stability and magnetic properties of  free standing 2D Ni nanostructures as a function of shape and lateral size, as shown in Figure \ref{fig:7}; namely two hexagonal flakes of increasing size (labeled {\bf H1} and  {\bf H2}), one triangular  island ({\bf T}), and the periodically extended 2D layer ({\bf L}) assumed as limiting reference. The main structural and magnetic parameters of these Ni clusters are summarized in Table \ref{tab:1}.

\begin{table}[htb]
\caption{Lateral size ($D$), lattice parameter (a$_0$), and average magnetic moment per atom ($\mu$) of selected finite and extended structures. The structure labeled 2H1 correspond to a bilayer of h1, whereas 2L denotes a bilayer stacking of L.}
\label{tab:1}
\begin{tabular}{|c|c|c|c|}
\hline
                  & $D$ (pm) & a$_0$ (pm) & $\mu$ (Bohr mag/atom) \\ \hline\hline
\textbf{H1}       & 136      & -          & 0.90                  \\ 
\textbf{H2}       & 318      & -          & 0.84                  \\ 
\textbf{T}        & 321      & -          & 0.88                  \\ 
\textbf{L}        & -        & 245        & 0.85                  \\ \hline
\textbf{2H1}      & 318      & -          & 0.76                  \\ 
\textbf{2L}       & -        & 245        & 0.68                  \\ \hline
\textbf{Gr/H1/Gr} & -        & 247       & 0.81                  \\ 
\textbf{Gr/L/Gr}  & -        & 247        & 0.79                  \\ \hline
\end{tabular}
\end{table}

After full atomic relaxation, all structures remain flat, with no significant distortions at the edges. All systems are metallic and magnetic, with an average magnetic moment $\mu\sim 0.9$ Bohr mag/atom (Table  \ref{tab:1}), uniformly distributed across the nanostructures, with no relevant differences between the center and the edges of the structures. Both shape and lateral dimensions have negligible effect on the electronic and magnetic properties of the systems.

Moving from left to right in Figure \ref{fig:7}, the density of states (DOS) spectra just become smoother, indicating an enhancement of the density of states for larger structures. The possible formation of Ni bilayers would not change this picture, as demonstrated for limiting cases of the {\bf 2H1} and {\bf 2L} dimers (Figure S11 of SI). In both cases, the vertical stacking followed the alignment of the Ni surface along the [111] direction, where Ni planes are separated by 197 pm; a distance that remained almost unchanged after the atomic relaxation and well below the distance of graphene layers. The interlayer interaction does not alter the main  electronic and magnetic properties of these Ni nanostructures, which maintained their metallic and magnetic character, albeit slightly reduced (Table \ref{tab:1}).

\begin{figure}[htb]
    \begin{center}
    \includegraphics[width=0.8\textwidth]{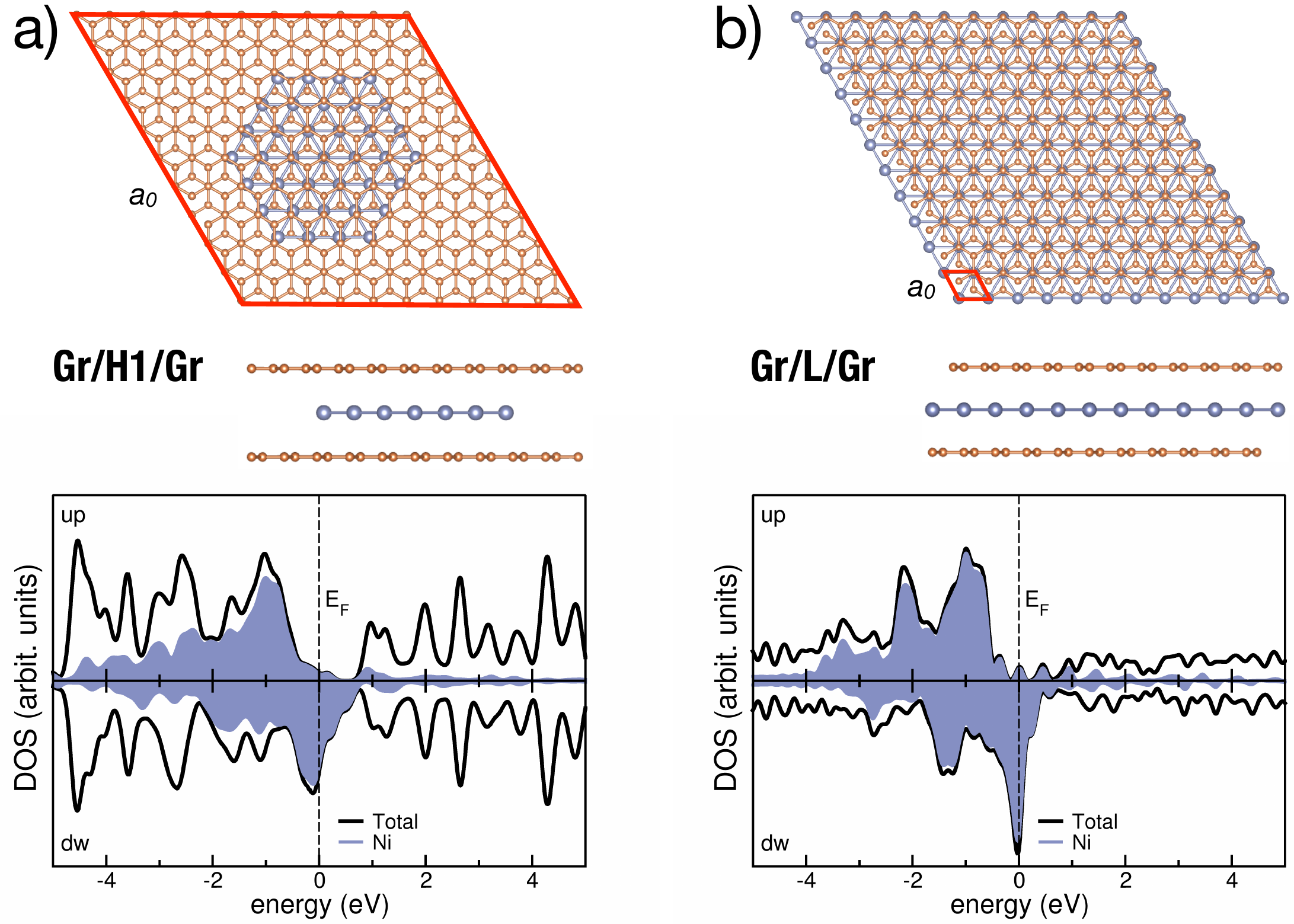}
    \end{center}
    \caption{Top and side view of atomic structure (top panel) and spin-resolved density of states (bottom panel) of 2D Ni nanostructures intercalated between graphene (Gr) bilayers: a) finite hexagonal Ni nanostructure {\bf Gr/H1/Gr}, and b) periodically repeated Ni layer {\bf Gr/L/Gr}. Labels refer to Figure 7. }
    \label{fig:8}
\end{figure}

Then, we consider the case of 2D Ni sandwiched between two extended graphene (Gr) layers. Since size and dimensions have a minor effect on Ni structures, we consider two stacks including the smallest nano\-structure (namely {\bf Gr/H1 / Gr}) and the extended layer (namely {\bf Gr/L / Gr}), as shown in Figure \ref{fig:8}. The initial models were built including Ni nanostructures within a bilayer of graphene.   For the {\bf Gr/H1/Gr} configuration (Figure \ref{fig:8}a), we considered a lateral periodicity (a$_0$) that corresponds to a $10\times10$ graphene structure. While for the  {\bf Gr/L/Gr} configuration (Figure \ref{fig:8}b), the simulation cell fits to the $1\times1$ graphene structure, yielding a C/Ni ratio of 2:1 per layer. In this case, the lateral periodicity of the Ni layer is slightly stretched ($\Delta a_0=+0.8\%$) to match the graphene lattice parameter. In both systems, the initial  vertical Ni-Gr distance was set to  230 pm, consistent with to the height of an intercalated island, measured in the experiments. After full atomic relaxation, Gr layers slightly detach increasing the Ni-Gr distance to 310 pm, while no significant relaxation takes place in both Ni and graphene structures. The corresponding ground state electronic structure is summarized in the bottom panels of Figure \ref{fig:8}. The resulting DOS spectra are almost the superposition of those of the single Ni and graphene layers. This result is consistent with the STM evidence, which indicates that the graphene lattice and its characteristic electronic properties remain largely intact above the intercalated Ni islands. The projection of Ni orbital states (shaded area)  closely reproduces the free standing ones of Figure \ref{fig:7}, as confirmed also by the similar magnetization (Table \ref{tab:1}). The slight reduction  of the average magnetic moment is an indication of Ni-Gr interaction. Yet, the magnetic reduction is lower than the one due to Ni-Ni stacking in the bilayer structures, indicating that the Ni-Gr coupling is weaker than the Ni-Ni one. The low Ni-Gr interaction is evident also from the the comparison between the band structures of the pristine graphene bilayer (i.e., without Ni interlayer) and the {\bf Gr/L/Gr} structure, as shown in Figure S12. Graphene bands remain clearly recognizable and spin unpolarized also in the presence of the  Ni layer, while the spin polarization of the stack completely derives from  Ni, which includes a d-like spin-polarized manifold of bands across the Fermi level.

\section{Conclusions}

In conclusion, this work provides an experimental demonstration of the detailed atomic and electronic structure of two-dimensional nickel stabilized at the buffer layer/monolayer graphene interface, as revealed by STM. The resulting intercalated 2D Ni nanostructures form small islands arranged in well-defined geometrical shapes, with edges oriented along the close-packed directions matching those of the $(6\sqrt{3} \times 6\sqrt{3})R30^\circ$ superlattice observed for monolayer graphene grown on SiC(0001). The observed stability of these structures under ambient conditions further enhances their technological appeal. Taken together, these findings position intercalated 2D Ni as a promising candidate for next-generation spintronic devices, opening new directions for the design of graphene-based magnetic heterostructures.

\section{Methods}

\subsection{\textit{Chemicals}}

Nickel(II) acetate tetrahydrate (98\% pure), cetyltrimethylammonium bromide (CTAB, $\geq$~98\% pure), and NaBH$_4$ ($>$ 98 \% pure) were purchased from Merck, while ethanol was supplied by Carlo Erba Reagent. Aqueous solutions were prepared using ultrapure water purified with a Millipore Milli-Q water system. All  reagents were used as received without further purification.

\subsection{\textit{Synthesis of Ni NPs}}

Ni NPs with a diameter of approximately 10 nm were prepared according to a procedure previously reported in  literature.\citep{Vlamidis2024}  Briefly, a solution of CTAB (12 mM) and  Ni(OAc)$_2$ (25 mM) in milliQ water was cooled to $\sim$1°C in an ice bath for 30 minutes. Then during vigorously stirring, 200 $\mu$L of sodium borohydride (20 mg/mL in milliQ water) was added in 5 aliquots, followed by an additional 2 minutes of stirring. After the metal reduction, the solution is centrifuged at 13000 rpm for 10 minutes to remove the precipitate containing large nanoparticles. The smaller nanoparticles (diameter $<$15 nm), suspended in the supernatant are collected by dilution of the solution with ethanol and washed several times with water by centrifugation and re-suspension, in order to remove excess reagents in the mixture.

\subsection{\textit{Epitaxial graphene growth}}

Epitaxial graphene was grown on the silicon-terminated face of 6H–SiC(0001) through thermal decomposition of the substrate at high temperature, under Ar atmosphere, in a Aixtron HT-Black Magic cold wall reactor.\citep{Emtsev2009} Prior to the growth, the samples were etched in hydrogen atmosphere to remove demages from chemo-mechanical polishing and to regularize the surface. 

\subsection{\textit{Sample preparation}}

Nickel NPs were dispersed in a 75:25 ethanol-to-water mixture, and the colloidal solution was sonicated for 10 minutes at room temperature to achieve a uniform distribution. The substrates were cleaned using isopropyl alcohol, dried under a nitrogen flow, and immersed in the colloidal solution (0.05 mg/mL) for several minutes before being left to dry.\citep{Vlamidis2024}

\subsection{\textit{Sample characterization}}

Scanning tunneling microscopy (STM)  measurements were carried out using an ultra-high vacuum RHK Technology STM system with a base pressure of 5$\cdot$10$^{-11}$ mbar. Tungsten tips were prepared by electrochemical etching in a 2 M NaOH solution, followed by in situ degassing and flashing to remove surface oxides. The samples were degassed at 200°C overnight and at 500°C for half an hour, and STM topographies were acquired after annealing the samples between 650 and 670 °C for durations between 2 and 10 minutes. The measurements were carried out in constant current mode with a tunneling current of 0.3-0.5 nA. Gwyddion software package was used to analyze the STM images.

X-ray Photoelectron Spectroscopy (XPS) measurements were performed using a Specs GmbH system equipped with an XR50 dual-anode source and a Specs Astraios 190 electron analyzer featuring a 60° angular acceptance and a 2D CMOS microchannel plate detector. Spectra were acquired using a non-monochromatized Al K$\alpha$ excitation source. EG samples were analyzed after functionalization with Ni NPs under as-prepared conditions and after annealing steps at 650 °C and 800 °C for 5 minutes.

Because of the geometry and low surface coverage of the nanoparticles, the contribution of Ni to the XPS signal is minimal. As a result, the Ni 2p signal remains weak even after prolonged acquisition times, allowing only the most prominent spectral features to be distinguished. Consequently, additional components such as shake-up satellites were not included in the spectral fitting.\\
Core-level spectra were fitted using an iterative Shirley-type background and symmetric Voigt line shapes (or doublets) for all components, except graphene and Ni, which were modeled using a Gaussian-broadened Doniach–Šunjić function with an asymmetry factor of 0.1.

\subsection{\textit{First Principles Simulations}}

Structural, electronic and magnetic properties of extended Ni films, nanostructures, and interfaces were simulated by using first principles approaches based on density functional theory (DFT), as implemented in the QuantumEspresso package \citep{qe2017}. Non-local van-der-Waals-corrected  formulation (vdw-df2-b86r) \citep{Hamada14} was assumed to describe the exchange-correlation functional.  Spin degrees of freedom are described within the local spin density approximation. Electron-ion interactions were described using norm-conserving pseudopotentials \citep{Hamann2013} from the  ONCVSP  library. Single-particle wavefunctions were expanded in planewaves with a kinetic energy cutoff of 70 Ry. A uniform mesh of  ($36\times36\times1$) $\mathbf{k}$-points was used to sample the 2D Brillouin zone  of extended films, while $\Gamma$ point is used in the case of finite nanostructures. Slab replicas were separated by $\sim15$~\AA{} of vacuum in the direction perpendicular to Ni planes, to avoid spurious interactions among replica. All structures were fully relaxed by using total-energy-and-force optimization scheme, until forces per atom are less than 0.03 eV/\AA.

\begin{acknowledgments}
The authors would like to acknowledge Andrea Camposeo and Francesca Matino for helpful discussions and for performing AFM analysis of a sample. The authors also acknowledge financial support from the PNRR MUR Project PE0000023–NQSTI, funded by the European Union – NextGenerationEU.
\end{acknowledgments}

\bibliography{Bib1}

\end{document}